\documentclass[twocolumn,amsmath,amssymb,longbibliography,superscriptaddress,10pt,prl,aps]{revtex4-1}

\usepackage{graphicx}
\usepackage{dcolumn}
\usepackage{bm}

\usepackage{amsfonts} 
\usepackage{mathtools}

\usepackage{esdiff}
\usepackage{braket}

\usepackage[english]{babel}

\usepackage[T1]{fontenc}
\usepackage[utf8]{inputenc} 
\usepackage{lmodern}
\usepackage{microtype}

\usepackage{color}

\usepackage{textcomp}

\usepackage[colorlinks=true,allcolors=blue]{hyperref}

\newcommand{\ii}{\mathrm{i}} 
 
\newcommand{\var}[1]{{(\Delta #1)^2}}
\newcommand{\rb}{$\prescript{87}{}{\text{Rb}}$ }

\begin{document}
\title{Quantum-Enhanced Sensing Based on Time Reversal of Nonlinear Dynamics}

\date{June 22, 2016}

\author{D. Linnemann}
\email{timereversal@matterwave.de}
\author{H. Strobel}
\author{W. Muessel}
\author{J. Schulz}
\affiliation{Kirchhoff-Institut f\"ur Physik, Universit\"at Heidelberg, Im Neuenheimer Feld 227, 69120 Heidelberg, Germany.}
\author{R. J. Lewis-Swan}
\affiliation{The University of Queensland, School of Mathematics and Physics, Brisbane, Queensland 4072, Australia.}
\author{K. V. Kheruntsyan}
\affiliation{The University of Queensland, School of Mathematics and Physics, Brisbane, Queensland 4072, Australia.}
\author{M. K. Oberthaler}
\affiliation{Kirchhoff-Institut f\"ur Physik, Universit\"at Heidelberg, Im Neuenheimer Feld 227, 69120 Heidelberg, Germany.}

\begin{abstract}

We experimentally demonstrate a nonlinear detection scheme exploiting time-reversal dynamics that disentangles continuous variable entangled states for feasible readout. Spin-exchange dynamics of Bose-Einstein condensates is used as the nonlinear mechanism which not only generates entangled states but can also be time reversed by controlled phase imprinting. For demonstration of a quantum-enhanced measurement we construct an active atom SU(1,1) interferometer, where entangled state preparation and nonlinear readout both consist of parametric amplification. This scheme is capable of exhausting the quantum resource by detecting solely mean atom numbers. Controlled nonlinear transformations widen the spectrum of useful entangled states for applied quantum technologies. 

\end{abstract}

\pacs{37.25.+k, 0.3.75.Gg, 0.3.75-b, 67.85.Fg, }

\maketitle

\begin{figure*}
\includegraphics{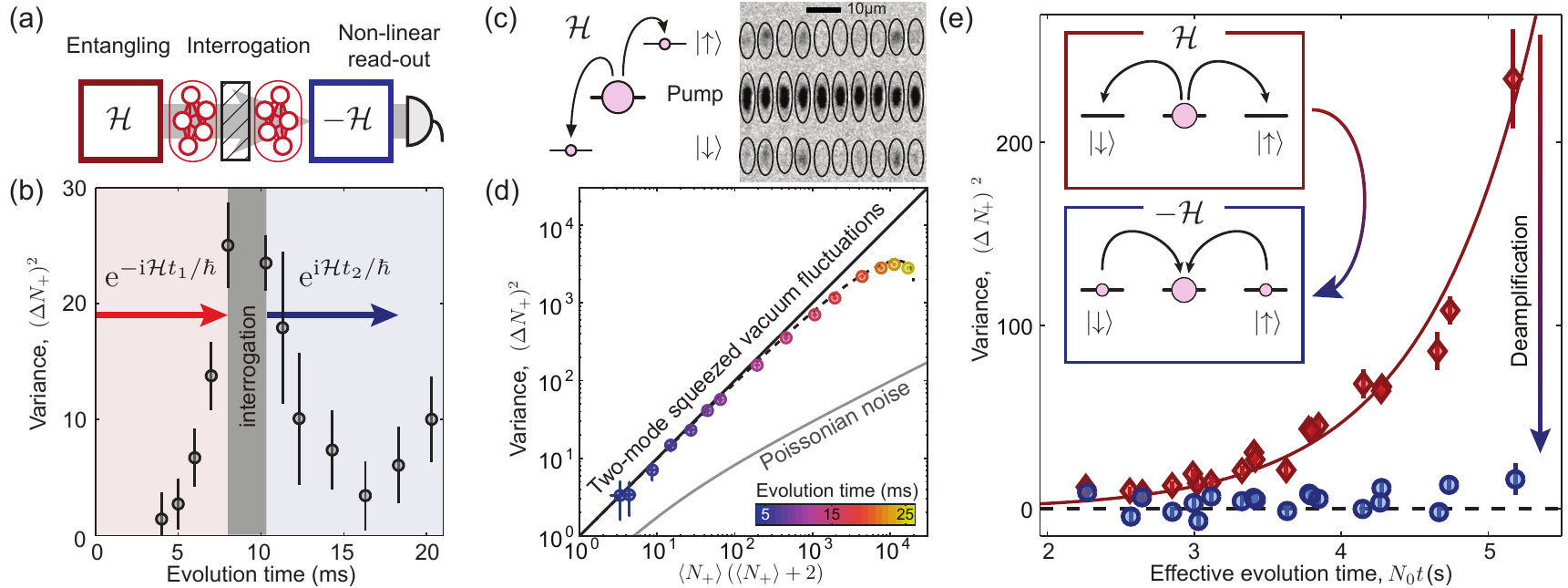}
\caption{Disentangling with nonlinear time reversal. (a) Our nonlinear readout scheme exploits a time reversal sequence. For this, the Hamiltonian $\mathcal{H}$ used for entangled state generation is inverted and reapplied for readout. Interrogation takes place in between both periods of nonlinear dynamics. (b) Time trace of the characteristic variance of $N_+=N_\uparrow+N_\downarrow$ during entangled state generation, interrogation, and time reversal. The initial drastic increase in variance is revoked by nonlinear evolution under the time inverted generation process. A pronounced minimum is found close to matched times, $t_1=t_2$. (c) Spin-changing collisions in a Bose-Einstein condensate are used as the nonlinear process. Atom numbers are detected by high resolution absorption imaging after Stern-Gerlach separation. A typical absorption image with counting regions indicated by ellipses is shown. (d) The side mode population exhibits characteristic thermal-like fluctuations, approaching the variance of the entangled two-mode squeezed vacuum state (diagonal). Results of a truncated Wigner simulation (dashed) and the expected variance of a coherent state (gray) are shown for comparison. (e) Variance of the side mode population before (red diamonds) and after (blue) time reversal sequence for the matched case of two equal durations of nonlinear dynamics. We find reversion to the initial vacuum state for a wide range of effective evolution times. The red line is a fit to the expected behavior in undepleted pump approximation.}
\end{figure*}

\begin{figure}
\includegraphics{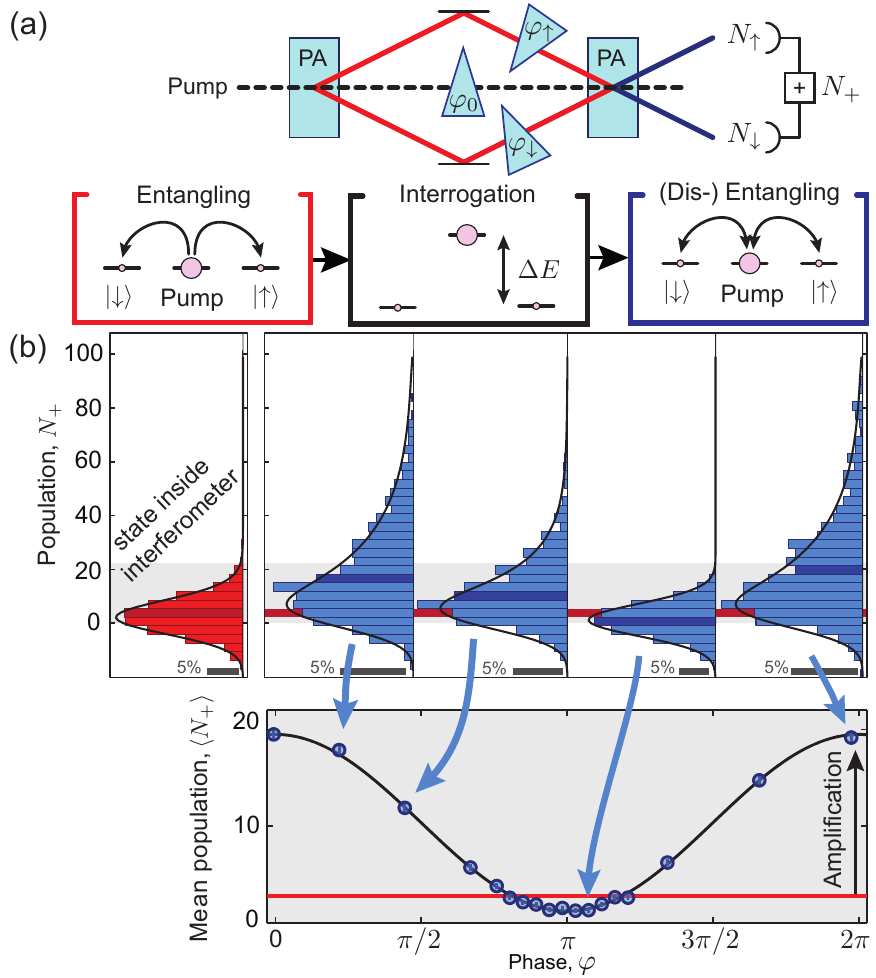}
\caption{Interferometry based on (dis-)~entangling nonlinear dynamics. (a) Schematic representation of an optical SU(1,1) interferometer and its realization in atom optics. This scheme takes advantage of the entanglement-enabled deamplification of fluctuations by time inversion of parametric amplification (PA). (b) Typical experimental population histograms of $N_+$ (black lines are fits to a thermal distribution convolved with our detection noise) for different spinor phase shifts $\varphi$ applied within the active interferometer. The blue histograms are recorded at the output, while the red one is obtained by omitting the final spin-mixing period. The dark colored bins depict the corresponding mean values, which are plotted in the lower panel (zoom-in into the gray shaded area), revealing the interferometry fringe. The horizontal dark red line denotes the average probe atom number of $\braket{N_{+}^\text{inside}}=2.8\pm0.2$ inside the interferometer.}
\end{figure}

Nonlinear dynamics is the basis of generating non-classical states of many particles. These entangled states are capable of improving a large variety of operations, e.g., computational tasks \cite{Ladd2010}, communication and measurements \cite{Giovannetti2006}. Unlocking their full potential for quantum technologies requires both the generation and detection at the fundamental quantum limit. The generation of such highly entangled states with many particles has witnessed tremendous advances \cite{Blatt2008, Pan2012}. However, to fully exploit this quantum resource, the complete correlations on the single particle level need to be accessed, which still limits current experiments. 

To address this challenge, nonlinear readout schemes have been proposed \cite{Dunningham2002,Leibfried2004,Froewis2015,Davis2016}. Most of these employ a time inversion sequence. For this the nonlinear evolution that is used to produce the entangled state is inverted and reapplied for readout. If the state remains unperturbed, the second period of nonlinear evolution counteracts the first. This time-reversed readout disentangles the probe state such that the known separable initial state is recovered. This reversibility is nonperfect if the state is changed in between, similar to an incomplete Loschmidt-Echo \cite{Goussev2012}. By this sensitive mechanism, minute state perturbations are mapped onto readily discernable quantities. 

Experimentally, we use spin-changing collisions \cite{Stamper-Kurn2013} in a mesoscopic spinor Bose-Einstein condensate. This nonlinear mechanism is the atomic analogue of parametric amplification, which is the textbook example of entangled state generation in quantum optics. At the same time, both the sign and the strength of the nonlinear coupling are experimentally adjustable, which makes this system ideally suited for realizing time reversal readout schemes.

Spin exchange is performed in an effective three-level system within the spin $F=2$ manifold of $^{87}$Rb. For this the external degrees of freedom are frozen out such that dynamics is restricted to the spin degree of freedom. We start with a pure $\ket{F=2,m_F=0}$ condensate (pump mode). Population in any $m_F\neq0$ state is carefully cleaned. During spin mixing atoms of the pump mode are coherently and pairwise scattered into the signal $\ket{\uparrow}\equiv\ket{2,1}$ and idler $\ket{\downarrow}\equiv\ket{2,-1}$ mode, which we refer to as side modes (see Fig. 1). For small population transfers from the highly populated pump mode, the spin-mixing dynamics is governed by the Hamiltonian $\mathcal{H} = \hbar \kappa \hat{a}_\uparrow^\dagger \hat{a}_\downarrow^\dagger + \text{h.c.}$, where $\hat{a}_\uparrow^\dagger$ ($\hat{a}_\downarrow^\dagger$) denotes the creation operator for the signal (idler) mode, $\hbar$ is the reduced Planck constant, and $\kappa$ is the effective nonlinear coupling strength. 

The coupling $\kappa=g N_0$ is related to the microscopic nonlinearity $g$, arising from coherent collisional interactions and is enhanced by the number of atoms $N_0$ in the pump mode. In this undepleted pump approximation, the pump mode is treated classically and serves as an unlimited particle resource for parametric amplification of the side modes which bears no dynamics of its own. We work within the physical $F=2$ manifold because its associated nonlinearity $g$ is one order of magnitude larger than for $F=1$. Spurious processes out of the effective three level system are energetically suppressed by the quadratic Zeeman shift at a magnetic field of $0.9\,$G.

The key feature of this three-mode implementation is that the nonlinear Hamiltonian can be tailored by controlling the phase and amplitude of this highly populated pump mode \cite{Hoang2013,Flurin2012,Vered2015}: The effective nonlinear coupling strength $\kappa$ is inverted by imprinting a phase shift of $2\varphi_0=\pi$, i.e., $\kappa\rightarrow\mathrm{e}^{-\mathrm{i}2\varphi_0}\kappa=-\kappa$, while its magnitude can be adjusted by the number of pump atoms.

We can therefore experimentally realize a scheme that is divided into three building blocks: Entangled state preparation, interrogation, and nonlinear time reversal for readout [Fig. 1(a)]. A characteristic quantity of the emerging entangled state is the fluctuation of $N_+=N_\uparrow+N_\downarrow$, where $N_\uparrow$ ($N_\downarrow$) denotes the amplified atom number in the respective mode. Figure 1(b) shows a measured time trace of the variance $\var{N_+}$ during this sequence. The independently characterized detection noise has been subtracted. During the first evolution (here up to $t_1=8\,$ms) it grows drastically indicating the generation of a highly entangled state. Following this, we allow for interrogation during which we inhibit spurious spin mixing. For this the pump atoms are transferred to $\ket{F=1,m_F=0}$ using a microwave $\pi$-pulse that is much faster than the spin exchange. Thereby the spin-mixing dynamics is halted and the pump is energetically shifted effectively by $100\,$Hz \cite{Note1}. We exploit this energy shift to imprint a dynamical phase of $2\varphi_0=\pi$ onto the pump mode that changes the sign of the spin-changing collisions Hamiltonian \cite{Hoang2013}. This phase accumulation takes $\sim 2\,$ms. We then rapidly transfer the pump atoms back to $\ket{F=2,m_F=0}$ and continue spin-changing collisions with identical coupling strength.  We find a pronounced minimum of the variance close to matched times of spin exchange, $t_1 \approx t_2$ as expected for this nonlinear time reversal sequence. The observed remaining variance in the minimum corresponds to $\sim 0.6$ atom per side mode on average. 

We now detail our first building block, which is the generation of the probe state. From a fundamental point of view, quantum-enhanced sensing relies on having entanglement at the probe stage - introducing entanglement solely after interrogation cannot yield increased sensitivity \cite{Giovannetti2006}. As we start spin-changing collisions with initially empty side modes, the process is analogous to optical parametric down-conversion, where amplifying vacuum fluctuations \cite{Leslie2009,Klempt2010} produces the paradigmatic two-mode squeezed vacuum state \cite{Gross2011, Hamley2012, Peise2015}. This entangled state is described by $\ket{\Psi} = \sum_{n=0}^\infty{\sqrt{p_n} \ket{n}_\uparrow\ket{n}_\downarrow}$, i.e. a coherent superposition of twin-Fock states. Within the undepleted pump approximation the weights $p_n$ are thermal-like, $p_n=\braket{N_\uparrow}^n/(1+\braket{N_\uparrow})^{n+1}$, where $\braket{N_{\uparrow}} =\braket{N_{\downarrow}}=\sinh^2(g N_0 t)$ is the mean atom number in either side mode after evolution time $t$. Because of the pairwise scattering during spin exchange, ideally both side modes are perfectly correlated, $N_-=N_\downarrow-N_\uparrow=0$. The side mode sum $N_+$, however, features distinctive excess number fluctuations with corresponding variances of $\var{N_+}=\braket{N_+}(\braket{N_+}+2)$ which are much larger than the Poissonian noise level $\braket{N_+}$.

To experimentally characterize this generated state and its broad distribution, we repeat the experiment typically a few thousand times. This is facilitated by simultaneously preparing up to $30$ independent condensates in a one-dimensional optical lattice potential. Atom numbers are detected via state and lattice site resolved absorption imaging with an uncertainty of $\pm4$ atoms \cite{Muessel2013}. A typical raw image is shown in Fig. 1(c). For quantitative analysis we postselect on total atom numbers in the range of $380-430$ atoms, corresponding to the initial pump population $N_0$, in order to restrict the nonlinear coupling strength $\kappa$.

We experimentally confirm in Fig. 1(d) that the variance of $N_+$ approaches the extreme value (solid black line) specific to the two-mode squeezed vacuum state. We find perfect agreement for evolution times up to $\approx 12\,$ms. For larger evolution times pump depletion limits the variance growth, such that the analogy to optical parametric amplification eventually does not hold any more. This effect is well captured by a numerical simulation based on the truncated Wigner method (dashed). 

Optimal reversibility of the spin-mixing process is achieved for short evolution times or small total atom numbers such that pump depletion is negligible. In Fig. 1(e), we systematically vary the spin-mixing nonlinearity by shifting the postselection window ($50\,$atoms) on total atom number and adjusting the evolution time $t$ of the spin exchange, which gives rise to an effective evolution time of $N_0 t$. After the nonlinear time reversal sequence (blue) we find good reversion to the initial vacuum state for a wide range of parameters. The red diamonds show the sum variance immediately before time reversal. 

The intrinsic phase dependence of the entangling Hamiltonian makes the entire scheme predestined for quantum-enhanced interferometry, where entangled states are employed to measure a phase shift more efficiently than classically allowed \cite{Giovannetti2006}. In linear interferometry with classical probe states, the precision in measuring a phase difference $\varphi_-$ between two modes $\ket{\uparrow}$ and $\ket{\downarrow}$ is bound by the standard quantum limit (SQL). The resulting phase sensitivity is given by $\var{\varphi_-}\geq\braket{N_+}^{-1}$, where $\braket{N_+}=\braket{N_\uparrow}+\braket{N_\downarrow}$ denotes the mean total atom number in both modes \cite{Pezze2015,Hofmann2009}. This limit can be overcome by exploiting the highly entangled two-mode squeezed vacuum as the input state, allowing phase estimation at the fundamental Heisenberg limit, $\var{\varphi_-}=[\braket{N_+}(\braket{N_+}+2)]^{-1}$ \cite{Pezze2015,Hofmann2009}. This precision can be reached by measuring the parity \cite{Anisimov2010} that necessitates single particle resolution. When having access to the global mean value only, a phase-dependent signal cannot be retrieved at all. 

In this work, we demonstrate that by using a nonlinear readout, the quantum resource can be harnessed by analyzing merely mean values. While number fluctuations of $N_-$ are necessary for a probe state to be sensitive to a phase difference $\varphi_-$ \cite{Giovannetti2006}, sensitivity to $\varphi_\uparrow+\varphi_\downarrow$ requires fluctuations of $N_+$, as inherent to the two-mode squeezed vacuum state. 

Phase accumulation during interrogation changes the probe state according to $\ket{\Psi} = \sum_{n=0}^\infty{\sqrt{p_n} \mathrm{e}^{\mathrm{i} n \left(\varphi_\uparrow+\varphi_\downarrow\right)} \ket{n}_\uparrow\ket{n}_\downarrow}$. This can be captured by modified mode operators for the consecutive spin-changing collisions period: $a_\uparrow^\dagger \rightarrow \mathrm{e}^{\mathrm{i}\varphi_\uparrow} a_\uparrow^\dagger$ (and similarly for $a_\downarrow^\dagger$). Thus, the second Hamiltonian evolution is characterized by a nonlinear coupling strength $\kappa\rightarrow\mathrm{e}^{-\mathrm{i}\varphi}\kappa$ where $\varphi=2\varphi_0-(\varphi_\uparrow+\varphi_\downarrow)$ is called the spinor phase. The initial evolution is reversed, $\mathcal{H}\rightarrow-\mathcal{H}$ if the well-controlled phase shift $\varphi_0$ of the pump mode satisfies $2\varphi_0=\pi+\varphi_\uparrow+\varphi_\downarrow$. Therefore, by determining $\varphi_0$ for which full time reversibility is reached, the unknown phase $\varphi_\uparrow+\varphi_\downarrow$ can be determined. 

The general phase dependence of $\braket{N_+}$ is the basis of the so-called SU(1,1) interferometer \cite{Yurke1986}. This has been proposed in the framework of nonlinear optics where it was realized recently \cite{Hudelist2014} with a bright seed in one side mode. Here, using an atomic system \cite{Marino2012} [see Fig. 2(a)] we explore the regime of unseeded side modes leading to maximally entangled probe states \cite{LewisSwan2013}. 

To characterize the phase dependence we continuously change the pump phase in between two equal periods of spin-changing collisions ($7\,$ms each). The probe state inside the interferometer is accessed by omitting the final spin mixing. Its atom number distribution is shown in Fig. 2(b) (red) featuring a mean atom number of $\braket{N_{+}^\text{inside}}=2.8\pm0.2$. All stated errors are statistical and are one s.d. Without accumulated phase ($\varphi \approx 0$), the interferometer's output corresponds to an overall spin mixing for twice the initial period, with the population of the probe state inside the interferometer being further amplified by a factor of $\approx7$. Compared to the ideal amplification factor of $2(\braket{N_{+}^\text{inside}}+2) \approx 9.5$, this is reduced due to pump depletion. 
For phases close to $\varphi=\pi$, time reversal yields a state with strongly reduced mean atom number. Each phase setting features the expected broad, non-Gaussian number distribution representative of the two-mode squeezed vacuum state. Remarkably, the corresponding mean values (dark colored bins) give rise to an interferometry fringe (lower panel). 

The nonlinear time reversal maps the phase information onto the collective quantity $\braket{N_+}$. Thus, the phase sensitivity of the entire device can be accessed by Gaussian error propagation given by $\var{\varphi} = \var{N_+}/|d\braket{N_+}/d\varphi|^2$ where only readily obtainable quantities enter.

\begin{figure}
\includegraphics{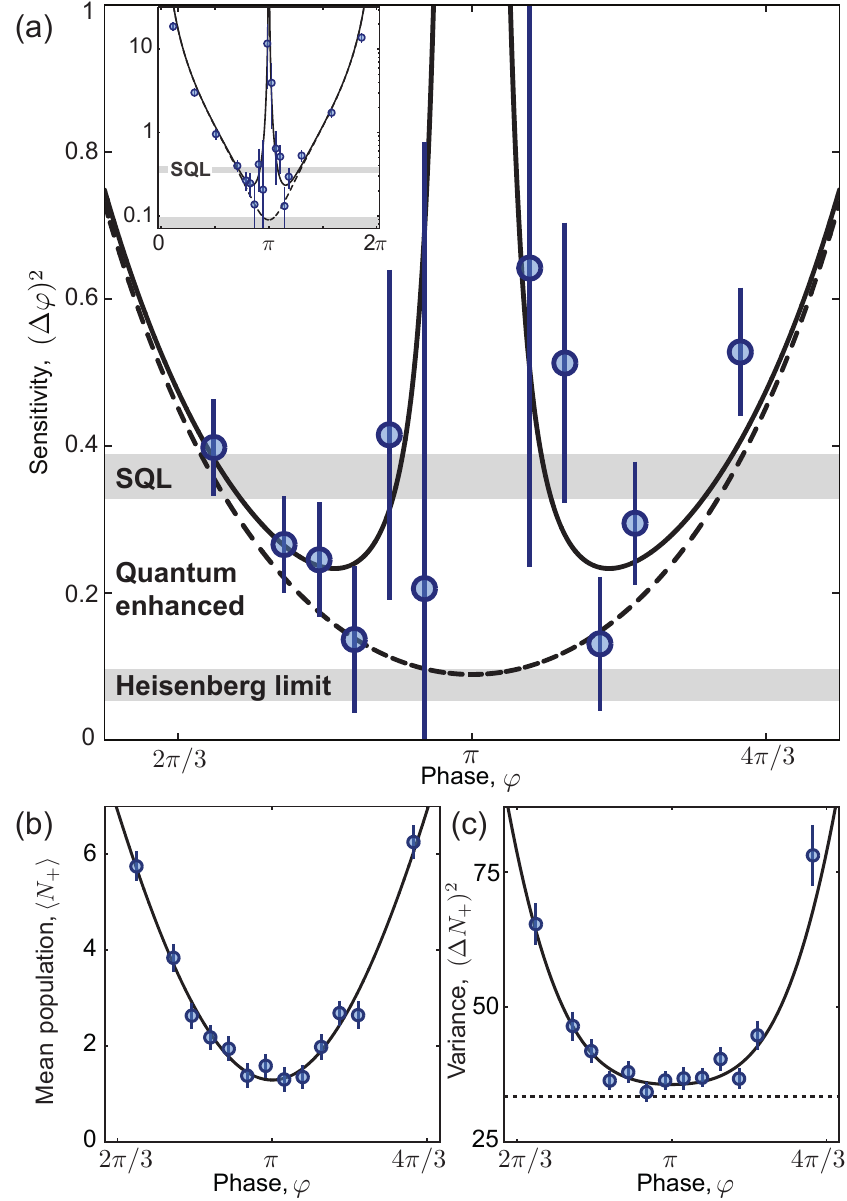}
\caption{Quantum-enhanced phase sensitivity with nonlinear readout. (a) The phase sensitivity is experimentally extracted by Gaussian error propagation on $\braket{N_+}$. The standard quantum limit (gray horizontal bar) is surpassed in close vicinity of phase $\varphi=\pi$. At phase $\pi$ the sensitivity diverges due to the vanishing slope of the signal. The undepleted-pump theory (dashed) additionally taking into account the nonperfect reversibility is shown as a solid line. The observed phase sensitivity agrees with the theoretical model of an active SU(1,1) interferometer over two orders of magnitude (inset).
(b) Mean side mode population $\braket{N_+}$ in vicinity of the fringe minimum. The signal's derivative is determined by the sinusoidal fit. (c) Variance of $N_+$ at the interferometer output. Our detection noise of $33.5\pm1.3$ is indicated by the horizontal dotted line and subtracted for determining the phase sensitivity. Good agreement to the undepleted-pump theory is found when considering the nonperfect reversibility by including an offset (black line).}
\end{figure}

Our experimental result is shown in Fig. 3(a): Over the full range of phases (inset) we find good agreement to the analytical undepleted-pump theory (dashed line). Specifically at the most sensitive working point of the SU(1,1) interferometer, the fringe minimum, a phase sensitivity at the Heisenberg limit is predicted \cite{Yurke1986, Marino2012, Gabbrielli2015} (dashed line). This is a consequence of both the increased slope of the signal due to the intrinsic amplification and the deamplified quantum-correlated noise at the minimum \cite{Kong2013,Pooser2009}. 

We determine the slope of the signal $d\braket{N_+}/d\varphi$ by a sinusoidal fit [solid line Fig. 3(b)] to the interferometer's output in close vicinity to the fringe minimum. By this we avoid underestimating the slope due to pump depletion, which affects only the maximum of the fringe. For the phase sensitivity [Fig. 3(a)] at the fringe minimum, a diverging signal is experimentally inevitable since nonperfect reversibility implies nonvanishing noise but zero slope of the signal. Nevertheless, we find the optimal regime with quantum-enhanced performance in close vicinity of the fringe minimum. The standard quantum limit $\var{\varphi}=\braket{N_+^\text{inside}}^{-1}$ \cite{Yurke1986, Hudelist2014}, and the corresponding Heisenberg limit $\var{\varphi}=[\braket{N_+^\text{inside}}(\braket{N_+^\text{inside}}+2)]^{-1}$, are determined by directly measuring the mean side mode population inside the SU(1,1) interferometer. 

Our observed variance is shown in Fig. 3(c) and reveals the expected shape within the undepleted pump approximation, characterized by a flattened variance around its minimum. We find quantitative agreement when taking into account the nonperfect reversibility by including a variance offset of a two-mode squeezed vacuum state with mean occupation number of $0.65\pm0.05$ atoms per mode on top of the independently characterized detection noise (dotted line). This occupation number is consistent with the observed minimum in panel (b), which suggests that nonideal reversibility rather than technical noise limits the performance. To infer the phase sensitivity, only the detection noise is subtracted, leading to the data points and the solid line in panel (a). Ideally, for reaching larger absolute phase sensitivities at the fringe minimum the side mode population inside the interferometer can be increased as long as the nonlinear Hamiltonian remains reversible which is strictly true only within the undepleted pump limit. 

Our findings point towards a new direction of accessing nonclassical resources for quantum metrology, employing highly controlled nonlinear dynamics for readout. Specifically the aspect that entanglement generated by nonlinear dynamics is best readout by time reversal \cite{Macri2016} opens up a new class of entangled states to be experimentally accessible even in the many-particle limit of strongly correlated quantum systems. We envision the time reversal as a modular and powerful tool for entangled state characterization and exploitation in the continuous variable regime, where efficient linear detection schemes remain challenging. 

\begin{acknowledgments}
We thank Luca Pezz\`e for insightful discussions. This work was supported by the Heidelberg Center for Quantum Dynamics, the European Commission small or medium-scale focused research project QIBEC (Quantum Interferometry with Bose-Einstein condensates, Contract No. 284584), the European Comission FET-Proactive grant AQuS (Project No. 640800), and the Go8/DAAD Australia-Germany Joint Research Co-operation Scheme. W.M. acknowledges support from the Studienstiftung des deutschen Volkes. K.V.K. acknowledges support by the Australian Research Council Future Fellowship Grant No. FT100100285.
\end{acknowledgments}

\textbf{Published in} Phys. Rev. Lett. {\bf 117}, 013001 (2016)\\
doi: \href{https://doi.org/10.1103/PhysRevLett.117.013001}{10.1103/PhysRevLett.117.013001}

\clearpage
\newpage

\renewcommand{\thefigure}{S1}

\section{Supplementary Material}

\section{Experimental setup}
Our experiments start with a Bose-Einstein condensate of \rb in the $\ket{F=1,m_F=-1}$ hyperfine ground state at a magnetic field of $B=0.9\,$G. It is trapped in a one-dimensional optical lattice ($5.5\,$\textmu m spacing, $\omega_l=2\pi\times660\,$Hz) superimposed with a harmonic trap ($\omega_t=2\pi\times440\,$Hz) for transversal confinement. The individual lattice sites contain $200$--$500$ atoms, tightly confined such that the dynamics happens in the internal degree of freedom. Since tunneling is negligible, the $30$ populated lattice sites are independent and used to increase the statistical sample size. 

\section{State preparation}
We transfer the atoms from the initial state to $\ket{1,0}$ by two resonant microwave ($\approx 6.8\,$GHz) $\pi$-pulses. Spurious atoms in $m_F\neq 0$ states are expelled by a strong magnetic field gradient at reduced depth of the optical potential. We then transfer the pure $\ket{1,0}$ condensate to $\ket{2,0}$ by a fast ($46\,$\textmu s) microwave $\pi$-pulse. This $\pi$-pulse is also used within the experimental sequence for ``shelving'' the pump atoms. 

\section{Detection}
After the experimental sequence we transfer the pump atoms from $\ket{2,0}$ to $\ket{1,0}$ to switch off the nonlinear coupling. State and lattice site resolved absorption imaging is used after Stern-Gerlach separation and $1\,$ms time of flight. The components $\ket{1,0}$ and $\ket{2,\pm1}$ are imaged simultaneously. The detection noise is determined by interleaved control measurements, where the atoms remain in $\ket{1,0}$ after $m_F$ cleaning. Extracting the background signal for each $m_F=\pm1$ component (same elliptical regions as in Fig. 1c) we find a Gaussian distribution of width $\sigma\approx4$ atoms centered at $\approx 0.3\,$atoms. The background offset might be caused by a slight tilt of the magnetic field direction between the Stern-Gerlach pulses used for cleaning and analysis and is subtracted for all data in the manuscript.

\begin{figure*}
\includegraphics[width=17.8cm]{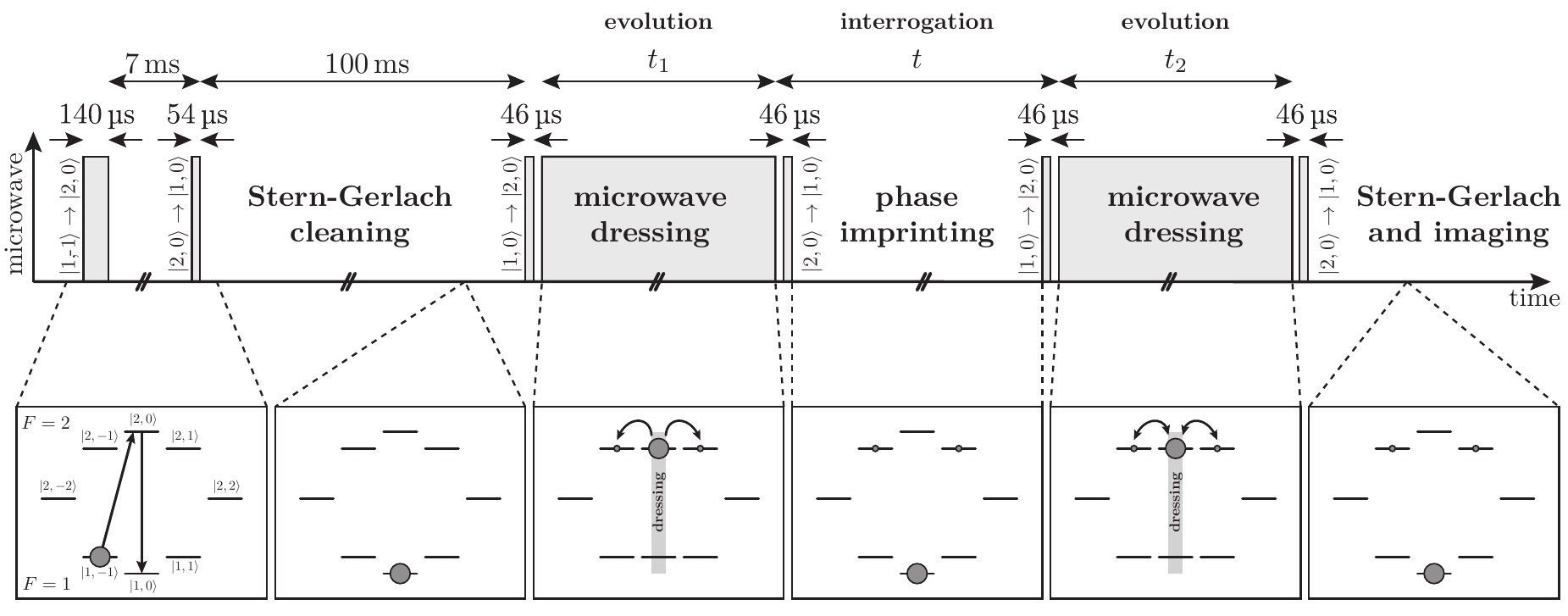}
\caption{Timing diagram. Microwave pulses used for state preparation and energy level shifting (dressing) are indicated in the upper panel. The lower panels show the energy levels at each stage with the linear Zeeman shift subtracted. }
\end{figure*}

\renewcommand{\thefigure}{S2}

\section{Hamiltonian}
The Hamiltonian of a spin-1 condensate in single spatial mode approximation can be split into three contributions, $\mathcal{H} = \mathcal{H}_\text{SCC} + \mathcal{H}_\text{el} + \mathcal{H}_B$ \cite{Law1998}. 

The spin-changing collisions are described by $\mathcal{H}_\text{SCC}=\hbar g (\hat{a}_0\hat{a}_0\hat{a}_\downarrow^\dagger \hat{a}_\uparrow^\dagger + \text{h.c.}$). For short evolution times the large pump mode remains undepleted and its operator $\hat{a}_0$ can be substituted by the c-number $\sqrt{N_0} \mathrm{e}^{-\ii \vartheta}$ with $\vartheta$ being the pump phase. We choose $\vartheta=0$ in the main text. Then $\mathcal{H}_\text{SCC}$ takes the form of parametric amplification. The second term, $\mathcal{H}_\text{el}=\hbar g(N_0-1/2)(N_\uparrow+N_\downarrow)$ describes the collisional shifts arising due to s-wave scattering of the three involved modes. $\mathcal{H}_B = \hbar q B^2 (N_\uparrow+N_\downarrow)$ with $q=2\pi\times72\,\text{Hz}/G^2$ contains the magnetic field dependence: Spin-changing collisions are magnetically insensitive to first order and only affected by the quadratic Zeeman effect that shifts the states of $F=2 \choose F=1$ according to $\Delta E={+\choose -} (4-m_F^2) \hbar q B^2$.

\section{Effective spin-1 system}
We use an effective three-level system within the physical $F=2$ manifold, encompassing $m_F = 0, \pm1$, which is effectively described by the spin-$1$ Hamiltonian. Its associated coupling strength $gN_0\approx2\pi\times20\,$Hz is one order of magnitude larger than for $F=1$. 
Spurious processes out of the three-level system, e. g. $2\times\ket{2,\pm1}\leftrightarrow\ket{2,\pm2}+\ket{2,0}$ or $2\times\ket{2,0}\leftrightarrow\ket{2,2}+\ket{2,-2}$ are energetically suppressed by the magnetic field shift and have a smaller coupling strength \cite{Widera2006}. 

The 1/e-lifetime of the large pump mode in $\ket{2,0}$ is $200\,$ms due to spin relaxation, while the small populations in the side modes have a lifetime exceeding $1\,$s. 

\section{Microwave dressing}
To fulfill the spin-mixing energy resonance $2\times\ket{2,0}\leftrightarrow\ket{2,1}+\ket{2,-1}$, i.e., to compensate the energy shift of $\mathcal{H}_\text{el}+\mathcal{H}_B$, we use microwave dressing \cite{Gerbier2006} $2\pi\times110\,$kHz blue detuned to the $\ket{1,0}\leftrightarrow\ket{2,0}$ transition. The resonant Rabi frequency of this magnetically insensitive transition is $\Omega=2\pi\times5\,$kHz and is stabilized by a power servo-loop. To precisely match the energies we record SU(1,1) interferometry fringes for both, different microwave dressing detunings and durations of spin exchange. These fringes are sensitive to the spinor phase accumulation during the spin-changing collisions, $\varphi_\text{pulse}$. Since the term $\mathcal{H}_\text{el}$ yields a dynamical phase mismatch (since it depends on the atom number in the side modes) its compensation works best for short times within the undepleted pump approximation. The microwave dressing is optimized such that this spinor phase accumulation stays small, $\varphi_\text{pulse}\approx10^\circ$, for the experimentally relevant evolution times of $6-10\,$ms. 

\section{Variance of the two-mode squeezed vacuum state}
Within the undepleted-pump approximation, each initially empty side mode population grows nonlinearly with corresponding number fluctuations of $\var{N_\downarrow}=\braket{N_\downarrow}(\braket{N_\downarrow}+1)$ and similarly for $N_\uparrow$. Due to the covariance of $N_\uparrow$ and $N_\downarrow$, the variance of $N_+$ is twice larger than the combined level of fluctuations, $\var{N_+} = 2(\var{N_\uparrow}+\var{N_\downarrow})=\braket{N_+}(\braket{N_+}+2)$. 
Error bars of variances are estimated by jackknife resampling.

\section{Phase imprinting} To efficiently halt spin mixing, we transfer the pump atoms to $\ket{1,0}$ and stop microwave dressing. Without this shelving of the pump, off-resonant spin mixing would continue in $F=2$ albeit microwave dressing is not applied. In contrast, in $F=1$ off-resonant spin mixing is negligible. 

During this time, the phase $2\varphi_0$ of the pump mode $\ket{1,0}$ evolves at a rate of $4qB^2\approx2\pi\times240\,$Hz compared to the side modes $\ket{2,\pm1}$ due to the magnetic field. The collisional shift of the pump, $gN_0\approx2\pi\times20\,$Hz, reduces this rate to $\omega=2\pi\times200\,$Hz.

After a holding time of $0-5\,$ms the pump is transferred back to $\ket{2,0}$ for the second spin-mixing period. 

\section{Phase sensitivity} Within the undepleted-pump approximation the fringe is given by $\braket{N_+}= V (1+\cos\varphi)$ where $V = \braket{N_+^\text{inside}}(\braket{N_+^\text{inside}}+2)$. Here, the phase is $\varphi=\omega t+2\varphi_\text{pulse}$ with $t$ being the interrogation time. The associated variance is flattened around the dark fringe: $\var{N_+}=2V(1+\cos\varphi)+[V(1+\cos\varphi)]^2$. 
The expected phase sensitivity is given by $\var{\varphi} = \var{N_+}/|d\braket{N_+}/d\varphi|^2 =\frac{1}{1-\cos{\varphi}}\left\lbrack \frac{2}{V}+(1+\cos{\varphi}) \right\rbrack$ \cite{Marino2012}.

\section{Comparison to linear detection schemes} 

\begin{figure}
\includegraphics[width=8cm]{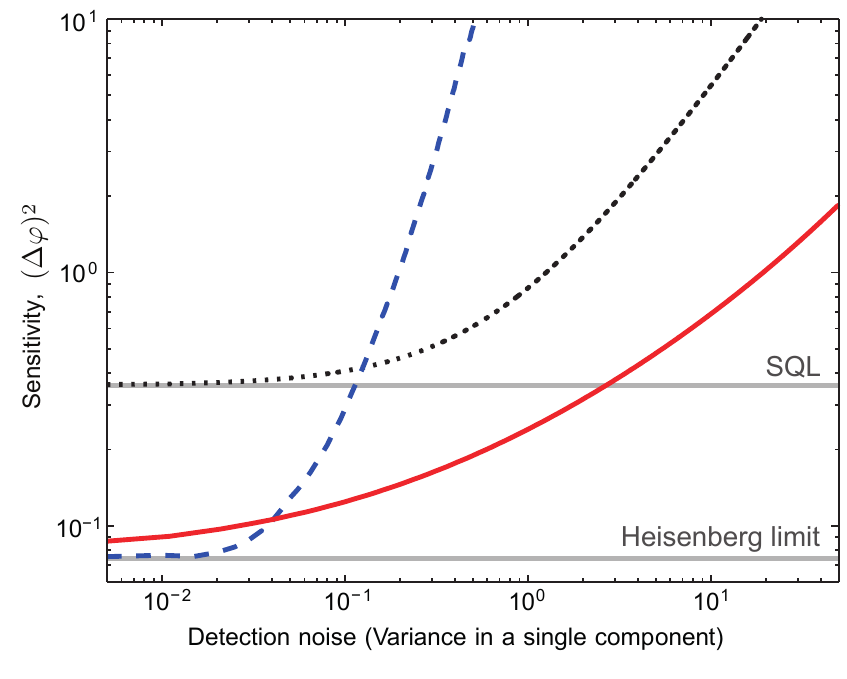}
\caption{Dependence of the phase sensitivity on detection infidelities for different interferometry schemes. A probe state with mean atom number $2.8$ is used in all cases. The two gray horizontal lines show the corresponding SQL and Heisenberg limit, respectively. The phase sensitivity of a Ramsey-sequence using a classical probe state is shown in black (dotted). Using the entangled two-mode squeezed vacuum state at its input allows reaching the Heisenberg limit. However, for linear detection the parity needs to be detected, which is highly susceptible to detection noise (blue dashed). The nonlinear detection strategy employed in this manuscript is shown as the solid red line. It features high phase sensitivities while being robust towards detection noise.}
\end{figure}

In Fig. S2 we compare the phase sensitivities of different interferometry scenarios when imperfect detection is taken into account. For this, a probe state with the same mean atom number of $2.8$ as in the manuscript is considered.

When using the two-mode squeezed vacuum state as the input for a Ramsey sequence, the Heisenberg limit (lower horizontal gray line) can be attained by parity detection \cite{Anisimov2010}. However, already small detection deficiencies reduce the sensitivity appreciably (blue dashed line). Detection noise of $\sigma^2=0.1\,$ atoms$^2$ reduces the single-shot sensitivity from the Heisenberg limit to the Standard Quantum Limit (SQL, upper gray line). For this calculation the ideal number distribution of one component is convolved with a Gaussian of variance $\sigma^2$.
The black dotted line shows the outcome of a Ramsey sequence employing a classical coherent state as probe. Here, the readout is the mean atom number difference which is robust against detection noise. However, because no entangled probe is used, the phase sensitivity is poorer than the SQL.
Our nonlinear detection scheme (red solid line) maps the phase-dependence of the highly entangled probe state onto the mean atom number. It therefore allows reaching high phase sensitivities below the SQL even in the presence of detection noise.

\section{Truncated Wigner simulation}
We model the spin-changing collisions of the effective three-level model using the truncated Wigner method (TWM) \cite{Gardiner_1985,Drummond_1993,Steel_1998}. The pump mode is represented initially by a coherent state, while the side modes are taken to be initially vacuum. Two-body loss is incorporated in the TWM, and we use loss coefficients extracted from experimental relaxation lifetimes \cite{Tojo_2009}. The parameters of the Hamiltonian are determined by a fit to the observed time evolution of $\braket{N_+}$.

\end{document}